# Robustness of an Artificial Intelligence Solution for Diagnosis of Normal Chest X-Rays

Tom Dyer,[1]⋆ Jordan Smith,[1] Gaetan Dissez, [1] Nicole Tay,[1] Qaiser Malik,[1]
Tom Naunton Morgan,[1] Paul Williams,[1] Liliana Garcia-Mondragon,[1] George Pearse,[1]
Simon Rasalingham[1]

[1] Behold.ai, 180 Borough High St, London SE1 1LB, UK

**Abstract**
**Purpose:** Artificial intelligence (AI) solutions for medical diagnosis require thorough evaluation to demonstrate that performance is maintained for all patient sub-groups and to ensure that proposed improvements in care will be delivered equitably. This study evaluates the robustness of an AI solution for the diagnosis of normal chest X-rays (CXRs) by comparing performance across multiple patient and environmental subgroups, as well as comparing AI errors with those made by human experts.
**Methods:** A total of 4,060 CXRs were sampled to represent a diverse dataset of NHS patients and care settings. Ground-truth labels were assigned by a 3-radiologist panel. AI performance was evaluated against assigned labels and sub-groups analysis was conducted against patient age and sex, as well as CXR view, modality, device manufacturer and hospital site.
**Results:** The AI solution was able to remove 18.5% of the dataset by classification as High Confidence Normal (HCN). This was associated with a negative predictive value (NPV) of 96.0%, compared to 89.1% for diagnosis of normal scans by radiologists. In all AI false negative (FN) cases, a radiologist was found to have also made the same error when compared to final ground-truth labels. Subgroup analysis showed no statistically significant variations in AI performance, whilst reduced normal classification was observed in data from some hospital sites.
**Conclusion:** We show the AI solution could provide meaningful workload savings by diagnosis of 18.5% of scans as HCN with a superior NPV to human readers. The AI solution is shown to perform well across patient subgroups and error cases were shown to be subjective or subtle in nature.

**Key words:** Artificial Intelligence – Radiology – Diagnostics

## 1 Introduction

Artificial intelligence solutions in healthcare have shown great promise to improve diagnostic accuracy, drive efficiency and reduce clinical workloads (1). However, many solutions do not perform as advertised when deployed to environments that are significantly different to the training and in-house testing environments (2,3). The challenge of building AI tools which are both generalizable in their performance and robust to imperceptible changes in input remains pertinent to the successful deployment across health systems (4).

In computer vision, many tools are developed using single-site datasets, or largely public datasets, collected from just a few institutions (5–8). This generates a risk that the resultant models will struggle to perform on patient populations that are 'unseen'. Models have been shown to learn hospital-specific configurations and artefacts (9), or perpetuate biases experienced by certain groups in a given institution (10). The risk of AI algorithms learning to make decisions along demographic boundaries is emphasised by recent work, showing algorithms can predict patient race from CXR with high accuracy (11). The requirement for multi-site evaluations of AI tools has been highlighted by recent works (12) as well as regulatory bodies stating the need to analyse potential performance differences across patient demographics and subgroups (13).

Multiple strategies have been proposed for both developing robust algorithms and validating robustness prior to deployment in the real world (14). These include training on larger, more representative datasets (15) and modifying input training data to reduce the potential for deep learning algorithms to 'overfit' to specific cases (2). In addition, the quality and consistency of labelled data has been shown to be important in preventing models learning hidden patterns in labelled data during training (16). Whilst label definitions and decisions made at the early stages of product development by machine learning practitioners can also impart significant bias at deployment time (17). The existence of unseen subsets of data where machine learning performance is meaningfully degraded have been highlighted as a significant issue in medical AI and dubbed 'hidden stratification' (18).

In addition to consistent performance across all subgroups of a given dataset, in safety-critical domains such as medical AI, robustness may also be interpreted as algorithms not making errors which would be obvious to a human expert (19). This 'justifiability' of error cases is of particular importance for diagnosis of medical imaging, where the consequences of errors is not constant and significant or obvious errors will likely impact on clinician and patient trust in AI systems.

⋆ corresponding author: tomd@behold.ai

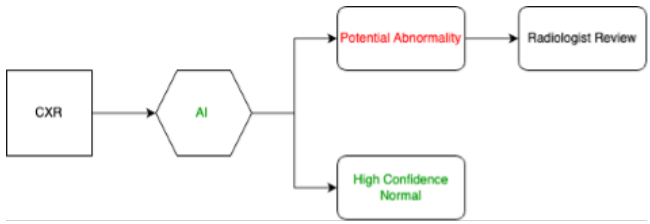

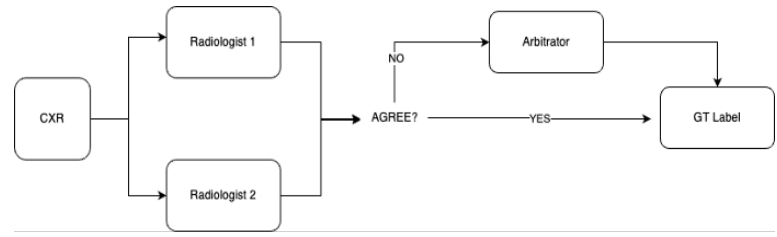

**Figure 1.** Schematic representation of rule-out of Normal scans by the AI solution, via classification of a subset of scans as 'High Confidence Normal' (HCN)

**Figure 2.** Process followed to assign ground truth labels to CXRs in the study.

In this study, we present a retrospective validation and robustness evaluation of an AI software solution for the diagnosis of normal chest X-rays (CXRs). The solution aims to detect a subset of the 'most normal' CXRs and classifies these as High Confidence Normal (HCN), whilst passing the remainder of scans to clinical experts for standard clinical review (Figure 1). This is proposed to provide a significant reduction in clinical workload, giving expert readers more time to review abnormal scans. First, we analyse AI performance against ground-truth labels and compare this to the performance of consultant-level radiologists, to determine whether the solution makes errors that human experts would not. Then, we present AI performance against multiple subsets of the dataset, broken down by patient subgroups such as age and sex, as well as the environment in which the scan was captured, including view, modality, device manufacturer and hospital site.

## 2 Materials and Methods

The proposed study is a retrospective validation of a deep learning solution for the detection of abnormalities and clinical rule-out of normal scans.

### 2.1 Study Dataset

The study dataset contained 4,060 adult CXRs collected from 4,060 unique patients. The scans were sourced from a total of 9 NHS hospital sites, all within England, with between 102 and 1,027 scans collected from each site (Table 2). The dataset was constructed to enable analysis of performance characteristics across metadata features such as view, modality, patient age, patient sex and device manufacturer. Due to this sampling, scans included were not collected consecutively and do not represent real-world disease distribution; this is discussed further in the limitations. All metadata features were collected from standard DICOM headers, with any patient identifying information being removed prior to analysis. As scans were collected retrospectively and anonymised, ethical approval was not required. Sub-optimal exams and AP films were not excluded from the study dataset, representing a challenging task for a rule-out normal AI tool.

### 2.2 Ground-Truth Labelling Protocol

Ground-truth labels were assigned by a pool of FRCR consultant radiologists with a minimum of 10 years' NHS experience. The ground-truth protocol (Figure 2) required each CXR to be initially read by two radiologists independently; if these reads

| Label | Count | % Dataset |
|---|---|---|
| Abnormal | 2258 | 55.6 |
| Sub-optimal Image | 582 | 14.3 |
| Cardiomegaly | 491 | 12.1 |
| Consolidation | 461 | 11.4 |
| Medical Device/Cosmetics | 454 | 11.2 |
| Nodule/Mass | 330 | 8.1 |
| Bony Abnormality | 252 | 6.2 |
| Fibrosis / Scar | 191 | 4.7 |
| Over Inflated Lungs | 168 | 4.1 |
| Pleural Effusion | 162 | 4 |
| Mediastinal Widening | 150 | 3.7 |
| Hilar Enlargement | 147 | 3.6 |
| Raised Hemidiaphragm | 147 | 3.6 |
| Interstitial Lung Shadowing | 125 | 3.1 |
| Atelectasis | 116 | 2.9 |
| Pleural Disease | 69 | 1.7 |
| Pulmonary Oedema | 59 | 1.5 |
| CCF | 53 | 1.3 |
| Pneumothorax | 30 | 0.7 |
| Collapse | 26 | 0.6 |

Table 1: The distribution of major disease labels as determined by the ground truth labelling protocol, for disease labels over 0.5% of the total dataset

agreed then ground-truth was assigned. However, if there was a discrepancy on any label, then a third radiologist would act as arbitrator and assign the final ground-truth label, with access to the findings of both radiologist 1 and 2.

The final ground-truth disease distribution can be found in Table 1. Reviewing radiologists could view the CXR pixel data, patient age and patient sex. No other contextual clinical information was provided to radiologists.

### 2.3 Definitions

In addition to the standard radiological findings reported in the study, the following definitions were provided to radiologists taking part in ground-truth labelling.

**Normal:**
*"A frontal image performed in inspiration showing a well-penetrated radiograph. Vertebrae are visible behind the heart. Left hemidiaphragm is visible to the edge of the spine. The lungs are appropriately visualised, and the vascular markings are not prominent. The entire chest*

*including lung apices and the costophrenic angles should be included in the field of view. Absence of abnormality in the lungs area, the mediastinum, pleural space, bones and upper abdomen. Absence of medical devices, except for electrocardiogram leads".*

**Sub-optimal:**
A sub-optimal image was defined as any image that did not satisfy all the below criteria:

- The entire chest including lung apices and the costophrenic angles is included in the Field of view
- No patient rotation
- Adequate depth of inspiration
- Adequate degree of photon penetration

## 2.4 AI Software

A commercially available UKCA Class IIa deep learning algorithm (red dot® v2.2, Behold.ai) was used to analyse all CXRs in the study. The device is designed to autonomously diagnose a subset of the most normal scans, designating these as 'High Confidence Normal' (HCN) and potentially significantly reducing radiological workload in reporting these scans. The software also has a functionality to prioritise lung cancer findings on CXR, which is not investigated in this study.

The software consists of an ensemble of convolutional neural networks, which has been trained on over 250,000 high-quality, labelled, and annotated CXRs. All abnormalities that do not conform to the provided definition of a Normal CXR should be identified by the software, which will output a continuous score between 0 (normal) and 1 (abnormal). The software processes pixel-level data only in its prediction of abnormality.

## 2.5 Radiologist Comparator

In order to provide context to the stated algorithmic performance in safely identifying normal scans, standalone radiologist performance was calculated. The initial independent reads by radiologists 1 and 2 in the ground-truthing process were used for this purpose (Figure 2). This has the limitation of being biased towards radiologist performance, as their label will have contributed to final ground-truth label. However, in the absence of further independent radiology reads, it provides an instructive perspective on human expert error rates in identifying abnormalities.

## 2.6 Hospital Sites

A total of 9 NHS hospital sites are included in the presented study. Although they are not named, the total number of scans provided, and geographical region of the trust is shown in Table 2. Fourteen scans had an unknown or 'other' hospital site entry and were excluded from analysis of performance across site of origin.

## 2.7 Statistical Analysis

Statistical performance of the AI software in identifying abnormalities was assessed against ground-truth labels. Of particular interest were false negative (FN) cases, i.e. ground-truth abnormal CXRs which were incorrectly classified as High Confidence Normal by the AI software, as this has a direct link to the clinical safety of the solution. Primary metrics used to assess this aspect were negative predictive value (NPV) and sensitivity for abnormal CXRs as well as other standard statistical metrics such as accuracy.

| Site | Region | CXRs included |
|------|--------|---------------|
| Site 1 | Southwest England | 1027 |
| Site 2 | Southeast England | 937 |
| Site 3 | East England | 476 |
| Site 4 | Southeast England | 400 |
| Site 5 | London | 316 |
| Site 6 | Midlands | 288 |
| Site 7 | Northwest England | 252 |
| Site 8 | South England | 248 |
| Site 9 | Southeast England | 102 |

Table 2: Breakdown of NHS hospital sites included in the study

For comparison of performance between dataset subgroups the Lakera software package (www.lakera.ai) was utilised in addition to custom python code. AUROC was used in addition to other standard metrics to account for differences in dataset distributions. Where appropriate, 95% confidence limits are calculated using bootstrapping methods utilising 10,000 iterations (20). The 95% confidence intervals calculated are then used to draw conclusions as to the statistical significance of differences between subgroups.

## 3 Results

Ground-truth labels were assigned for all CXRs in the study dataset. A total of 1,802 exams (44.4%) were labelled as normal. All CXRs were processed by the AI software with exams being classified as either High Confidence Normal (HCN) or not (i.e. potentially containing an abnormality).

### 3.1 AI Software Performance

Of the total 4,060 CXRs, the AI software classified a total of 751 exams as HCN, representing 18.5% of all scans and 41.7% of all available normal scans. Ground-truth labels showed that 723 scans were correctly identified as normal, giving an NPV of 96.0%. The sensitivity for abnormal scans was 98.7%, having missed 28 abnormalities that were identified by radiologists during the ground-truth process.

In each of the 28 abnormalities missed by the AI software, one radiologist initially classified the CXR as normal, but all were then arbitrated as containing an abnormality. This suggests all were subjective or subtle findings which could be missed in clinical practice. The full list of abnormalities missed by the software is shown in Figure 3. A selection of example cases of abnormalities missed by the AI solution are shown in Figure 4. These cases are selected based on the commonality of missed abnormality and potential clinical severity of the disease label. In all cases one radiologists labelled the scan as normal.

### 3.2 Radiologist Agreement and Performance

Across the entire dataset, a total of 1,864 CXRs required arbitration by a third radiologist, suggesting a total agreement rate of 54.1% between radiologists across all disease labels. The agreement rate between radiologists for the simpler nor-mal/abnormal classification was much higher, at 82.0%. Arbitration strongly favoured confirming the presence of abnormalities, with just 19.2% of the cases reviewed by a third radiologist being labelled as 'Normal'. In a total of 534 cases, one of the radiologists reported a CXR as normal when the other identified an abnormality which was later confirmed by the arbitrating radiologist. This corresponds to an average sensitivity of 88.2% for all abnormalities and a negative predictive value of 89.1%, significantly lower than the figures shown for the AI software. The disease label distribution of findings missed by at least one radiologist are shown in Figure 5.

| Patient Sex | Count | % Total | % Normal | % HCN | NPV | FN | Abnormal AUC |
|---|---|---|---|---|---|---|---|
| M | 1831 | 45.1 | 42.6 | 17.2 | 96.8 (94.4, 98.3) | 10 | 0.92 (0.91, 0.93) |
| F | 2211 | 54.5 | 45.8 | 19.4 | 95.8 (93.6, 97.4) | 18 | 0.92 (0.91, 0.93) |

Table 3: Performance breakdown of the AI solution against patient sex. 95% confidence intervals are given for NPV and Abnormal AUC.

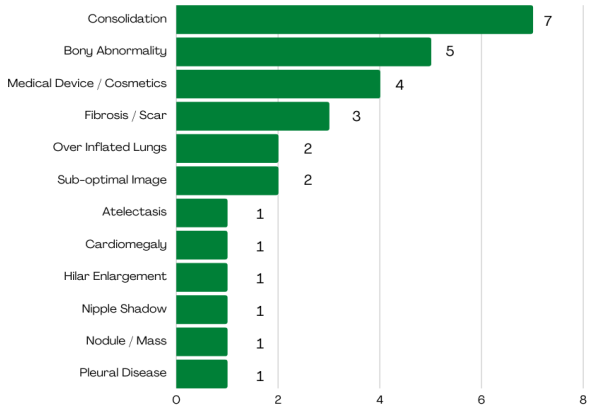

**Figure 3.** Breakdown of disease labels included in false negative cases from the AI software as determined by ground-truth labels. The total number of missed abnormalities is greater than the total number of false negative exams as some exams contain multiple abnormalities.

### 3.3 Performance Against Patient Sex

AI performance in identifying HCN scans for male and female patients is shown in Table 3. The solution showed an identical AUC score of 0.92 for abnormality detection and no significant variations in NPV were observed. A greater proportion of female patients overall were classified as HCN. In the study dataset, 18 cases had a patient sex entry of 'other' and were excluded from analysis.

### 3.4 Performance Against Patient Age

Performance on patients of different age groups is shown in Table 4. Notably, the proportion of scans classified as HCN decreases with increased age. This mirrors the overall decrease in the percentage of scans with ground-truth normal labels. This is likely to reflect a real increase in clinical abnormalities for advanced age groups, not simply age-related changes, as reporting radiologists were provided with patient age.

No statistically significant variations are observed between age groups in the study. Figure 6 shows how the proportion of scans classified as HCN by the AI solution declines with the overall proportion of normal cases.

### 3.5 Performance Against Scan View

AI performance across CXR view (PA/AP) can be seen in Table 5. As would be expected, a much lower proportion of AP scans were assigned ground-truth normal labels. The AI solution classified just 3.6% of AP scans as HCN but no false negative cases were recorded.

### 3.6 Performance Against Scan Modality

AI performance across CXR modality is shown in Table 6. No significant variations are observed between computed radiography and digital radiography modalities.

### 3.7 Performance Across Device Manufacturers

A total of 9 distinct device manufacturers were included in the study dataset. 14 scans had 'Other' recorded in the DICOM elements and were excluded from this section of the analysis. AI performance in classifying HCN scans across these manufacturers can be seen in Table 7. For multiple manufacturers, no false negative cases were recorded, so 95% confidence limits for NPV could not be calculated. For all other manufacturers, no significant variations in NPV were observed, although confidence intervals were broad for some manufacturers, owing to relatively small sample sizes.

### 3.8 Performance Across Hospital Site

CXRs were collected from 9 distinct NHS hospital sites for this study, details of each site can be found in the methods section (Table 2). The performance of the AI software in assigning HCN classifications for CXRs from each hospital site can be found in Table 8.

The proportion of scans labelled as normal varies significantly between sites, from 15% to 75%. Correspondingly, the percentage of scans classified as HCN by the AI algorithm varies from 3.5% to 30.6%. However, no significant variations are seen in NPV between hospital sites and less variation is observed in the proportion of available normal scans being removed as HCN, with the AI software removing between 23% and 56% of available normal scans at 8 of the 9 sites. The exception to this is site 9, where the AI software only classified 3.9% of scans as HCN, despite 42.16% of total CXRs being normal. However, no false negative cases were observed. This outlier may be due to the smaller sample size at this hospital site, with just 102 scans included in the analysis. A slightly lower portion of scans (3.5%) were classified as HCN from site 4, however the exams from this site had fewer normal scans (15.04%).

## 4 Discussion

This study presents further evidence for the utility of an AI software solution for the autonomous diagnosis of a subset of the most normal CXRs. The software was able to identify 723 out of 4,060 (18.5%) scans as High Confidence Normal (HCN) with just 28 false negative cases. This negative predictive value of 96.0% was superior to that of radiologists involved in the ground-truthing process, who displayed an average NPV of 89.1%. The potential to autonomously report these scans as described in Figure 1 could have a meaningful impact on radiological workload at a time when the Royal College of Radiologists forecasts a shortfall in the consultant radiologist workforce of 39% by 2026 (21). .

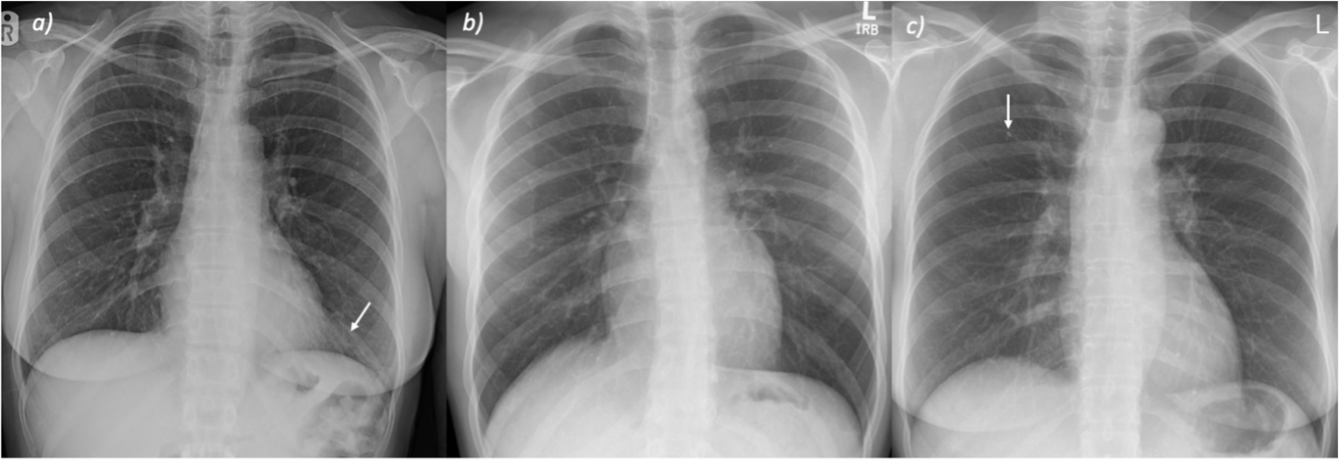

**Figure 4.** Example false negative CXRs, as classified by the AI software. A) subtle consolidation in the lower left lung, arrow indicates the area highlighted by the arbitrating radiologist. B) Chronic bony abnormality, scoliosis of the spine. C) Small nodule in the right lung, measuring 2mm in size, arrow indicates the area highlighted by the arbitrating radiologist. All CXRs were classified as normal by one radiologist. Full size images are available in the supplementary material.

| Patient Age | Count | % Total | % Normal | % HCN | NPV | FN | Abnormal AUC |
|---|---|---|---|---|---|---|---|
| 18 - 30 | 278 | 6.85 | 79.14 | 50.7 | 98.58 (95.1, 100) | 2 | 0.92 (0.88, 0.95) |
| 30 - 40 | 369 | 9.09 | 68.02 | 46.6 | 95.93 (91.9, 98.2) | 7 | 0.9 (0.86, 0.93) |
| 40 - 50 | 463 | 11.4 | 64.79 | 35.2 | 94.48 (90.1, 97.2) | 9 | 0.9 (0.86, 0.92) |
| 50 - 60 | 698 | 17.19 | 56.88 | 22.9 | 96.88 (93.0, 98.8) | 5 | 0.92 (0.9, 0.94) |
| 60 - 70 | 736 | 18.13 | 41.44 | 11 | 93.83 (86.3, 97.7) | 5 | 0.88 (0.86, 0.9) |
| 70 - 80 | 770 | 18.97 | 30.65 | 3.4 | 100 | 0 | 0.9 (0.88, 0.92) |
| 80 - 90 | 591 | 14.56 | 12.69 | 0.2 | 100 | 0 | 0.9 (0.86, 0.93) |
| 90 + | 142 | 3.5 | 9.15 | 0 | n/a | 0 | 0.89 (0.82, 0.94) |

Table 4: Performance breakdown of the AI solution against patient age group. 95% confidence intervals are given for NPV and Abnormal AUC. In cases where there are no false negative cases or no exams classified as HCN, confidence intervals are not included.

| View | Count | % Total | % Normal | Total HCN | % HCN | NPV | FN | Abnormal AUC |
|---|---|---|---|---|---|---|---|---|
| PA | 3452 | 85.02 | 48.9 | 723 | 20.9 | 96.13 (94.6, 97.4) | 28 | 0.92 (0.91, 0.93) |
| AP | 608 | 14.98 | 18.75 | 22 | 3.6 | 100 | 0 | 0.94 (0.91, 0.96) |

Table 5: Performance breakdown of the AI solution against CXR view.

Ground-truth labelling by the participating radiologists highlights the subjective nature of CXR interpretation, with just 54.1% agreement across all disease labels. In Figure 5, we show that a large number of potentially significant findings are reported as normal by the participating radiologists. A significant portion of radiologist false negatives are sub-optimal images (113), suggesting this may have represented a subjective disease classification. There are notably 59 cases of missed nodules or masses, which could represent clinically significant errors if these are later confirmed to be lung cancer. We hypothesise that reducing overall workload may enable greater

| Modality | Count | % Total | % Normal | Total HCN | % HCN | NPV | FN | Abnormal AUC |
|---|---|---|---|---|---|---|---|---|
| DX | 2587 | 63.72 | 47.2 | 488 | 18.9 | 96.72 (94.9, 98.1) | 16 | 0.92 (0.91, 0.93) |
| CR | 1472 | 36.26 | 39.47 | 257 | 17.5 | 95.33 (92.0, 97.5) | 12 | 0.93 (0.92, 0.94) |

Table 6: Performance breakdown of the AI solution against CXR modality.

| Manufacturer | Count | % Total | % Normal | % HCN | NPV | FN | Abnormal AUC |
|---|---|---|---|---|---|---|---|
| Agfa | 144 | 3.55 | 54.17 | 19.4 | 100 | 0 | 0.91 (0.85, 0.95) |
| Canon | 484 | 11.92 | 42.36 | 11.6 | 96.43 (87.3, 100) | 2 | 0.92 (0.9, 0.94) |
| Carestream | 450 | 11.08 | 56.89 | 26 | 100 | 0 | 0.92 (0.89, 0.94) |
| Fujifilm | 671 | 16.53 | 30.4 | 11.5 | 98.7 (94.2, 100) | 1 | 0.93 (0.9, 0.94) |
| GE | 311 | 7.66 | 66.56 | 19.9 | 98.39 (90.7, 100) | 1 | 0.88 (0.84, 0.92) |
| Kodak | 176 | 4.33 | 37.5 | 9.7 | 94.12 (71.7, 100) | 1 | 0.91 (0.85, 0.94) |
| Other | 14 | 0.34 | 42.86 | 14.3 | 100 | 0 | 0.98 (0.73, 1.0) |
| Philips | 208 | 5.12 | 44.23 | 18.3 | 92.11 (77.4, 97.6) | 3 | 0.9 (0.85, 0.93) |
| Samsung | 793 | 19.53 | 47.67 | 23.6 | 93.58 (89.0, 96.3) | 12 | 0.93 (0.91, 0.94) |
| Siemens | 809 | 19.93 | 38.32 | 19.9 | 95.03 (90.5, 97.6) | 8 | 0.95 (0.93, 0.96) |

Table 7: Performance breakdown of the AI solution against CXR device manufacturer.

| Hospital | Count | % Total | % Normal | % HCN | NPV | FN | Abnormal AUC |
|---|---|---|---|---|---|---|---|
| Site 1 | 1027 | 25.3 | 36.12 | 14.7 | 94.04 (88.6, 96.8) | 9 | 0.93 (0.92, 0.95) |
| Site 2 | 937 | 23.08 | 37.46 | 20.9 | 93.37 (89.5, 96.5) | 13 | 0.95 (0.93, 0.96) |
| Site 3 | 476 | 11.72 | 56.72 | 25.4 | 100 | 0 | 0.92 (0.89, 0.94) |
| Site 4 | 399 | 9.83 | 15.04 | 3.5 | 100 | 0 | 0.93 (0.89, 0.95) |
| Site 5 | 317 | 7.81 | 75.39 | 30.6 | 96.91 (92.0, 99.1) | 3 | 0.87 (0.82, 0.91) |
| Site 6 | 288 | 7.09 | 66.32 | 25.3 | 98.63 (93.8, 100) | 1 | 0.89 (0.84, 0.92) |
| Site 7 | 252 | 6.21 | 68.65 | 22.6 | 96.49 (87.9, 100) | 2 | 0.89 (0.84, 0.93) |
| Site 8 | 248 | 6.11 | 39.92 | 12.5 | 100 | 0 | 0.94 (0.91, 0.96) |
| Site 9 | 102 | 2.51 | 42.16 | 3.9 | 100 | 0 | 0.83 (0.73, 0.9) |

Table 8: Performance breakdown of the AI solution by hospital site.

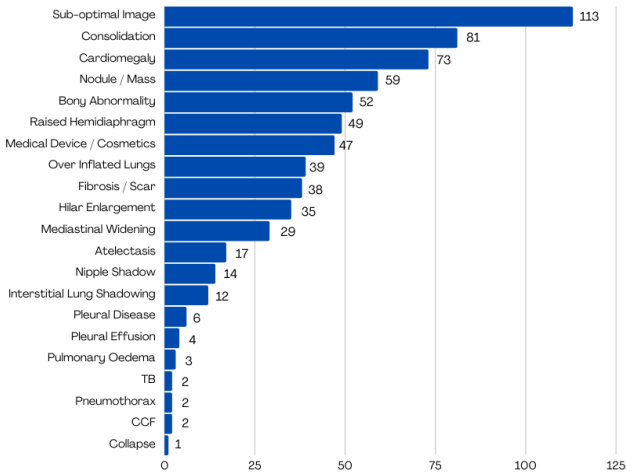

**Figure 5.** Breakdown of disease labels included in false negative cases from radiology reads as determined by the final, arbitrated ground-truth labels.

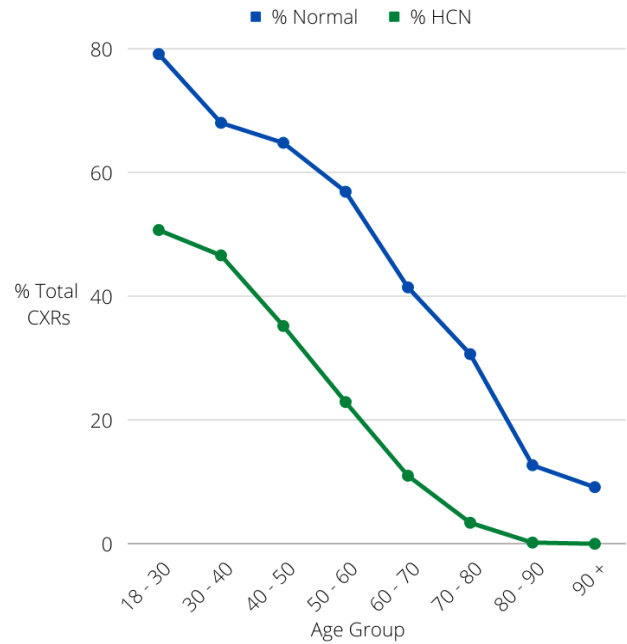

**Figure 6.** Percentage of all CXRs labelled as Normal by ground-truth labels and classified as High Confidence Normal by the AI solution, broken down by patient age group.

scrutinization of remaining exams, potentially reducing these missed findings.

The characterisation of AI errors is essential in building clinician and patient trust, particularly in autonomous tools (19) In this study, we demonstrate that the AI solution made no errors that were not also made by at least one radiologist, suggesting these scans contain subtle or subjective pathology. Whilst the clinical outcomes are beyond the scope of this study, this result indicates that integration of the AI tool is unlikely to lead to significant degradation in patient care on a diagnostic level, although further prospective investigations would be required to review the overall impact. Furthermore, the study dataset represented a challenging mix of common and rare pathologies, including many sub-optimal scans (14.3%). Performance on these cases provides supporting evidence that the AI solution is robust in the context of clinical deployment.

In this study, we have presented extensive evidence of AI performance across multiple hospital sites, device manufacturers, view, modality, age, and sex, with no significant variations observed in the safety of the AI solution on any subgroup. Some subgroups show reduced classification of HCN cases, potentially missing an opportunity to deliver further resource savings, however this is not at the expense of diagnostic safety. The importance of multi-site validations for clinical AI tools has been emphasised (12) and in this study we go further to validate performance across multiple other patient and environmental characteristics.

### 4.1 Limitations

Our study was designed to investigate the robustness of an AI solution across patient demographics and subgroups in a retrospective fashion. There are several limitations with this approach. Firstly, the dataset has been collected to optimise the balance of patient subgroups included in the analysis. This does however mean a deviation from standard NHS patient distributions, which may impact the strength of conclusions made regarding the ability of the algorithm to identify normal scans.

In addition to this, the retrospective nature of the study means that conclusions are limited to standard algorithm performance metrics. It is important that prospective assessments of AI solutions are performed, as it is possible patient subgroups will be impacted differently by the use of AI in their care beyond the accuracy of their diagnosis.

Whilst we have provided analysis across a broad spectrum of metadata features such as device manufacturer, this list is not exhaustive of all devices which may be present within the NHS or other healthcare markets. However, the presented data provides a good level of confidence that the algorithmic solution is robust to variations in device manufacturer. We also acknowledge that although we provide a high number of different hospital sites, certain regions such as Wales, Scotland, Northern Ireland, and the north of England are under-represented in our study.

Finally, whilst our study takes into account multiple patient and demographic features, factors like ethnicity and socio-economic status are not considered. These data points were not available as part of the study and collection was beyond the planned scope, however generalized performance across these features remains critical to the ethical deployment of AI solutions.

### 4.2 Conclusion

In this study, we show that the tested AI solution can rule out 18.5% of all scans with a negative predictive value of 96.0%, superior to the average radiologist NPV of 89.1%. In the 28 cases where abnormalities were missed by the algorithm, a radiologist had also classed the exam as normal, suggesting subtle or subjective pathology. No errors were made by the AI solution that were unanimously classified as abnormal by radiologists, indicating an increased acceptability of these error cases. The algorithm also shows a good level of robustness across a number of patient and environmental-level subgroups, indicating it would perform well in any NHS institution..

## 5 Disclosure

**Funding:** The authors state that this work has not received any funding

**Conflicts of interests:** GD, NT, TD, LM, JS, GP, SR are employed by Behold.ai. QM, TNM, PW and SR has stock/stock options in Behold.ai.

**Ethical Approval:** Institutional Review Board approval was not required because the study was conducted retrospectively. All data were anonymised prior to analysis.

# Supplementary Materials

*Supplementary Figure 1 – full-size versions of CXRs presented in Figure 4. The original annotation made by the arbitrating radiologist is included in each case (annotations were not required for scoliosis). A) Subtle consolidation in the lower left lung; arrow indicates the area highlighted by the arbitrating radiologist. B) Chronic bony abnormality, scoliosis of the spine. C) Small nodule in the right lung, measuring ~2mm in size; arrow indicates the area highlighted by the arbitrating radiologist.*

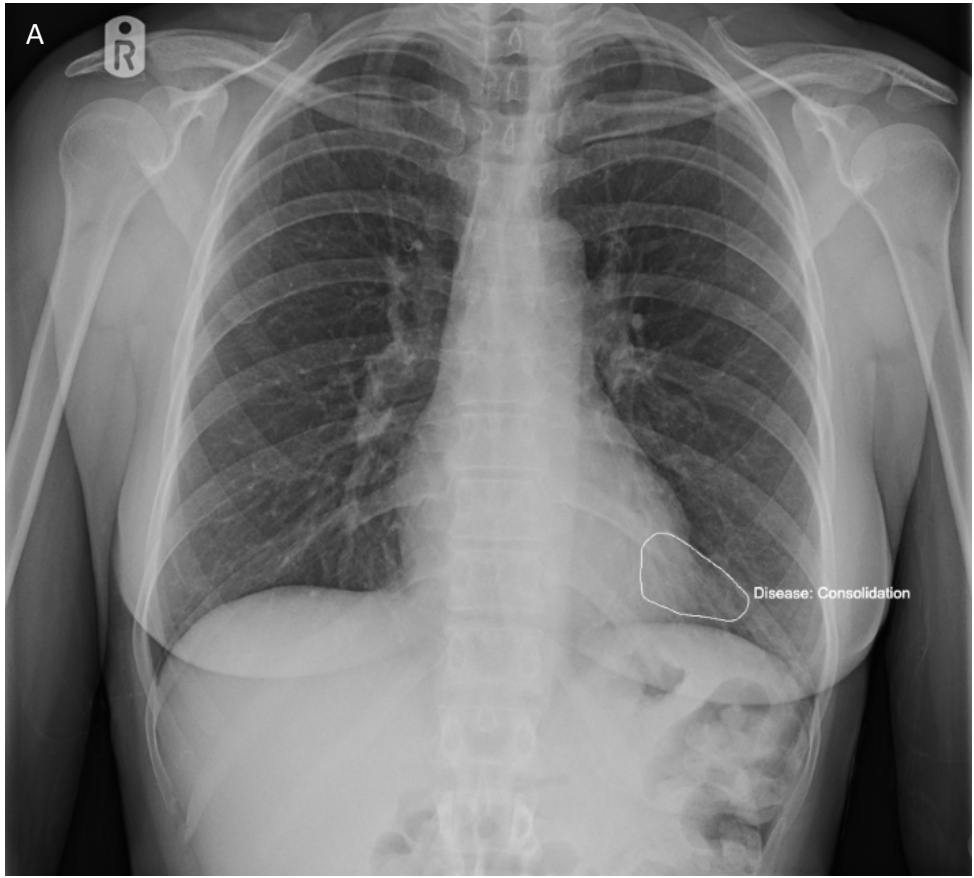

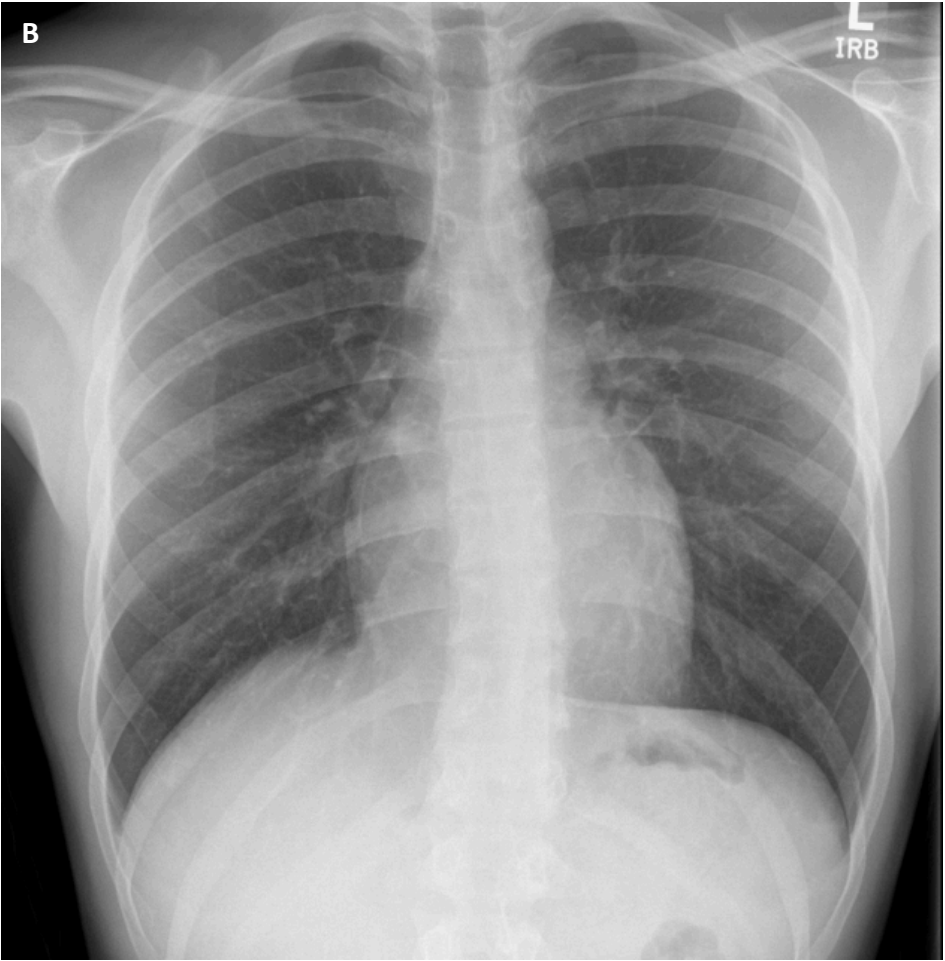

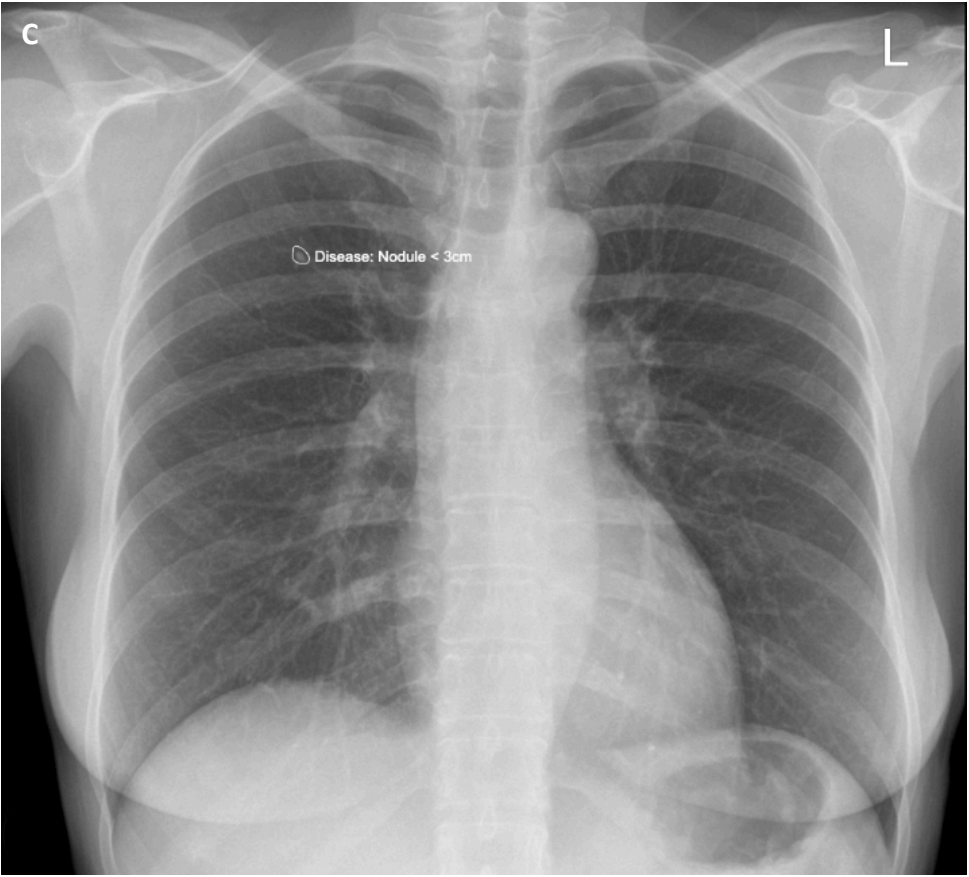